\documentclass[smallextended]{svjour3}

\usepackage{graphicx}
\usepackage[caption=false]{subfig}
\usepackage{multicol}

\usepackage{amssymb}

\usepackage{natbib}
\usepackage{url}

\usepackage{xspace}
\usepackage{float}

\usepackage{ifpdf}
\ifpdf
       \usepackage[pdftex,colorlinks,pdfauthor={Hugo Buddelmeijer}]{hyperref}
       \DeclareGraphicsExtensions{.pdf,.png,.jpg}
\else
       \usepackage[hypertex,colorlinks]{hyperref}
\fi

\usepackage{algorithm}
\usepackage{algorithmic}

\newcommand{\AW}{{\sf Astro-WISE}\xspace}

\newcommand{\refsec}[1]{section~\ref{#1}\xspace}
\newcommand{\reffig}[1]{Fig.~\ref{#1}\xspace}
\newcommand{\reftab}[1]{table~\ref{#1}\xspace}

\newcommand{\Reffig}[1]{Fig.~\ref{#1}\xspace}

\newcommand{\booleantrue}{\texttt{True}\xspace}
\newcommand{\booleanfalse}{\texttt{False}\xspace}
\newcommand{\setor}{\cup}
\newcommand{\setand}{\cap}

\newcommand{\vennheight}{0.08\linewidth}
\newcommand{\incvenn}[1]{\includegraphics[height=\vennheight]{hypercube#1}}

\begin{document}

\title{Leveraging Data Lineage to Infer Logical Relationships between Astronomical Catalogs}

\titlerunning{Inferring Logical Relations Between Catalogs}        

\author{
Hugo Buddelmeijer
           \and
 Edwin A. Valentijn
}

\institute{
Hugo Buddelmeijer \at
Kapteyn Astronomical Institute, Postbus 800, 9700 AV, Groningen, The Netherlands \\
\email{buddel@astro.rug.nl}
           \and
 Edwin A. Valentijn \at
\email{valentyn@astro.rug.nl}
}

\date{Received: date / Accepted: date}

\maketitle

\begin{abstract}
A novel method to infer logical relationships between sets is presented.
These sets can be any collection of elements, for example astronomical catalogs of celestial objects.
The method does not require the contents of the sets to be known explicitly.
It combines incomplete knowledge about the relationships between sets to infer a priori unknown relationships.
Relationships between sets are represented by sets of \textit{Boolean hypercubes}.
This leads to deductive reasoning by application of logical operators to these sets of hypercubes.
A pseudocode for an efficient implementation is described.

The method is used in the \AW information system to infer relationships between catalogs of astronomical objects.
These catalogs can be very large and, more importantly, their contents do not have to be available at all times.
Science products are stored in \AW with references to other science products from which they are derived, or their dependencies.
This creates full \textit{data lineage} that links every science product all the way back to the raw data.
Catalogs are created in a way that maximizes knowledge about their relationship with their dependencies.
The presented algorithm is used to determine which objects a catalog represents by leveraging this information.

\end{abstract}

\section{Introduction}
A set is a collection of elements.
For example, an astronomical catalog is a set with celestial objects as elements.
These sets have relationships with one another, for example a set could be a subset of another set.
The relationships between sets can be inferred by comparing their elements.
However, this is only possible when it is feasible to iterate over all the elements in the sets.
A novel method is presented that does not require the contents of the sets to be known explicitly.
A priori unknown relationships between sets are inferred by combining incomplete information that is available.

The method is designed for the \AW information system to infer relationships between astronomical catalogs \citep{pullingcatalogs}.
Catalog handling using this method is discussed in the following sections.
However, the method is generic enough to be used for other purposes.

Catalogs can be stored and even used in \AW without determining their full contents.
The creation of the catalog data is postponed until necessary and the result is only stored when beneficial for performance.
That is, the information system will only derive those parts of a catalog that are required for further processing.
As a result, the catalog data might not be available as a whole when the catalog is used.
One of the key aspects of \AW is that science products are automatically found or created when requested.
This requires the information system to be able to infer the contents of the catalogs automatically.
Determining the contents of the catalogs has to be possible without requiring access to the catalog data itself, since this might not be stored.

\AW stores science products with all the information required to (re)create the data.
In particular, every science product is stored with links to other science products from which it was derived, called its \textit{dependencies}.
This creates full \textit{data lineage} that links data products all the way back to the raw data.
As a result of this, every catalog `knows' from which other catalogs it is derived.
In particular, it is known which relations might hold between the sets of sources of a catalog and those of its dependencies.
A priori only this local information about the relationships between catalogs is available.
A more global overview of the relationships between catalogs is necessary for the desired automation.
The presented method combines this local information to achieve the required knowledge.

The novelty of the method is the use of Boolean hypercubes to represent relations between sets.
Relationships between specific sets are represented as sets of hypercubes in order to account for incomplete knowledge.
This makes it possible to deduce relationships by application of logical operators on these sets of hypercubes.

Ultimately, the presented method is a specialized form of automated theorem proving.
Other such methods could be used to infer relationships, for example software that can solve the problems in the SET domain of the TPTP Problem Library\footnote{\url{http://www.cs.miami.edu/~tptp/}}\citep{SutcliffTPTP}.
Those methods are very generic and can be used to solve several kinds of logical problems.
The presented mechanism is more specific:
while the used hypercube representation is natural for dealing with sets, it is not directly applicable to more general problems.

Relational databases can use similar mechanisms for query optimization (see for example \citet{Chaudhuri:1998:OQO:275487.275492} for an overview).
These are embedded in the optimization algorithms and are therefore not directly applicable to the requirements of \AW.

This paper is structured as follows.
The representation of relationships by means of sets of hypercubes and the details of the method are given in \refsec{sec:descriptionofalgos}.
Applications of the presented mechanisms in \AW are discussed in \refsec{sec:astrowise}.
Subsequently, the pseudocode of the algorithms is given in \refsec{sec:pseudocode}.
This is followed by an example in \refsec{sec:example2} and conclusions in \refsec{sec:conclusions}.

\section{Description of Algorithms}
\label{sec:descriptionofalgos}
The basis of the method is the use of Boolean hypercubes to represent logical relations between sets (\refsec{sec:relashyper}). 
The relationships between specific sets are represented by means of a set of hypercubes to account for incomplete knowledge (\refsec{sec:relasset}).
Deduction is possible through application of logical operators on the sets of hypercubes (\refsec{sec:logicaloperationsonrels}).
Scalability in implementation is achieved by optimizing important logical operators (\refsec{sec:optimizations}).
Pseudo code for an implementation of the method is given in \refsec{sec:pseudocode}.

\subsection{Relations as Hypercubes}
\label{sec:relashyper}
\begin{figure}[ht!]
    \centering
    \subfloat[Hypercube]{\label{fig:hypercuberep}\includegraphics[width=0.25\linewidth]{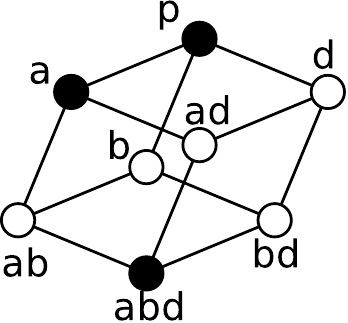}}
    \hspace{1cm}
    \subfloat[Array]{\label{fig:matrixrep}\includegraphics[width=0.25\linewidth]{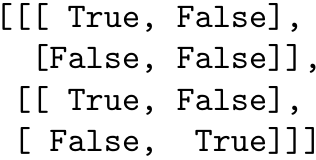}}
    \hspace{1cm}
    \subfloat[Shaded Venn]{\label{fig:vennrep}\includegraphics[width=0.25\linewidth]{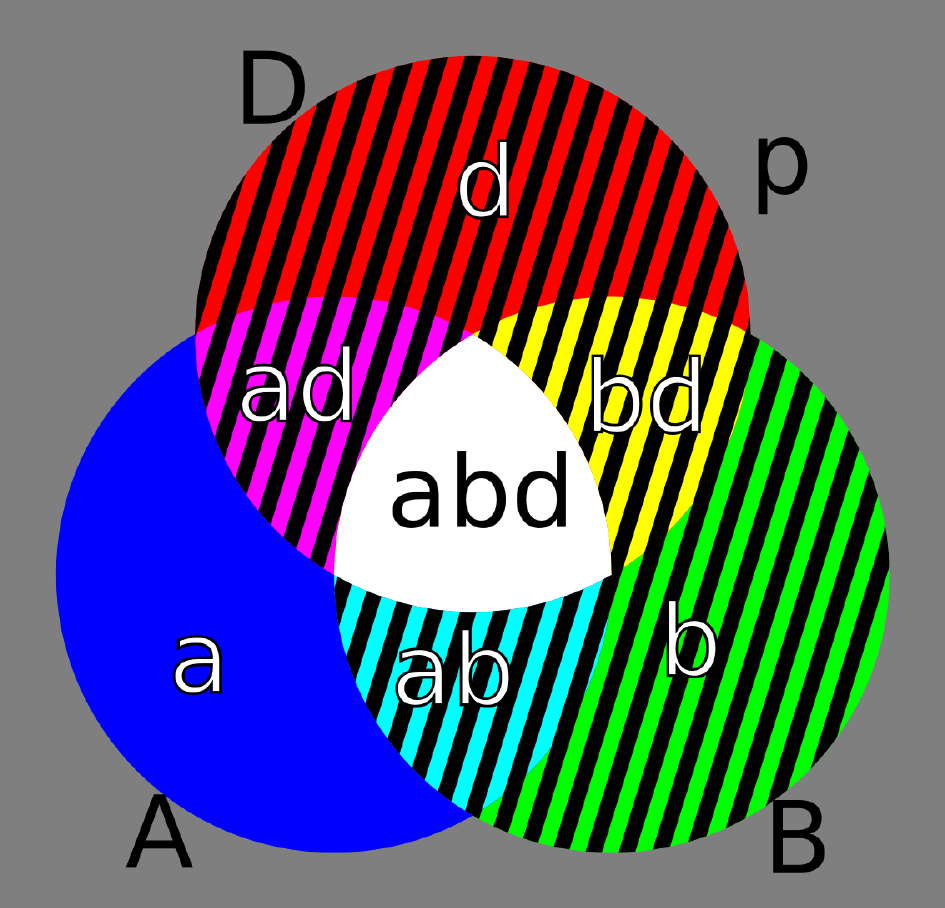}}
    \caption{Hypercube, array and shaded Venn representation of the relationship between sets $A$, $B$ and $D$.
      The vertices represent intersections between sets as indicated with lower case letters.
      The vertex labeled $p$ represents objects not in any set under consideration.
      Solid vertices indicate that the corresponding intersection is non-empty.
      The array representation can be used verbatim in the Python programming language by means of the numpy package.
      None of the sets are empty, set $B$ and $D$ are equal and set $A$ is a superset of them, but does not contain all objects in the universe.
    }
    \label{fig:matrixhypercuberep}
\end{figure}

A Boolean hypercube can be used to represent a relation between sets by associating a set to each dimension of the hypercube (\reffig{fig:hypercuberep}).
This representation is well suited for a numerical implementation by means of a multidimensional array (\reffig{fig:matrixrep}). 
Examples of hypercube representations of relations are given in \reftab{tab:hypercubeslowd}.
In particular the relations that are directly relevant to our astronomical application are shown.

Every logical relation between $n$ sets can be represented by means of an $n$-di\-men\-s\-i\-o\-nal hypercube.
This is done by identifying each of the $2^n$ possible intersections between the sets with one of the vertices of the hypercube.
A vertex in the second position of a specific dimension represents objects that are elements of the set corresponding to that dimension.
A vertex in the first position of the dimension represents objects not in the corresponding set.
For example, the vertex that is in the second position in all dimensions represents objects that are in all the sets described by the hypercube.
The vertex that is in the first position in all dimensions represents objects that are in none of the sets under consideration.
A Boolean value can be assigned to each vertex to indicate whether the corresponding intersection between sets contains any objects:
a Boolean \booleantrue value is assigned if the vertex represents one or more objects and a Boolean \booleanfalse value is assigned if it does not.
The collection of all objects---whether inside a set or not---is called the \textit{universe}, which can be empty.

This hypercube representation of relations between sets is similar to Karnaugh maps \citep{k:map:1953} and to the hypercube representation of logical operators by \citet{Clarke199497short}.
Furthermore, the hypercubes can be translated into shaded Venn diagrams \citep{Venn1880} by assigning every vertex to a region of overlap in the Venn diagram.

A hypercube of a certain dimension also represents specific relations of lower dimensions.
A lower dimensional relation is inferred by summing the hypercube over the dimensions that represent the unwanted sets (Algorithm \ref{algo:removeset}).
Summing over a dimension means repeatedly performing the logical \textit{or} operator on two adjacent vertices that are aligned in that dimension, since Boolean values are assigned to the vertices.

\subsection{Relationships as Sets of Hypercubes}
\label{sec:relasset}
A relationship between specific sets is described with a set of all hypercubes that are consistent with the available knowledge about the relationship.
This stems from our astronomical requirements, where the exact relationship between sets is not always known.
For example, there are four different hypercubes that represent an equality: between empty or nonempty sets and with or without objects outside the considered sets (\reftab{tab:hypercubeslowd}).
Representing that two sets are identical, without any extra available information, should therefore be done with a set of these four hypercubes.
However, more information is usually not necessary: it is enough to determine that the relationship between two sets must be one of these four in order to infer that they are equal.

The set of hypercubes representation allows us to define four classes of relationships between sets:
\begin{itemize}
 \item The \textit{Contradiction}, an empty set of hypercubes: there is no relation between the sets that is consistent with the available knowledge.
 \item An \textit{Exact Relation}, a set with exactly one hypercube: there is only one relationship possible between the sets; everything is known about the sets under consideration.
 \item An \textit{Inexact Relation}, a set with more than one hypercube: there are several relations that are consistent with the available knowledge.
 \item The \textit{Tautology}, a set with all $2^{2^n}$ possible hypercubes representing $n$ sets: every relationship is possible; nothing is known about these sets.
\end{itemize}
The use of the tautology can be prevented in a numerical implementation (\refsec{sec:optimizations}).
It is included here because it is useful in discussing the presented mechanisms.

A relationship also represents knowledge about sets that do not correspond to any dimension of the hypercubes.
For example, an empty universe can be represented with a hypercube of any dimension with \booleanfalse as the value of all vertices.
Such a relation implies that all sets, also those that have no corresponding dimension in the hypercube, must be empty.
Most relationships are less strict: in general they tend to represent the tautology for sets that have no corresponding dimension.

A higher dimensional relationship can be inferred from a lower dimensional one by adding dimensions to the hypercubes.
This results in a set of hypercubes that can be constructed as follows (Algorithm \ref{algo:addsetsimple}):
first create the tautology for the higher dimension.
Subsequently remove all hypercubes that are not consistent with the original relationship.
Adding a set to a relationship usually results in an increase of the number of hypercubes necessary to represent the relationship.

\subsection{Logical Operations on Relations}
\label{sec:logicaloperationsonrels}
A natural way to apply logical operators to relations follows from the use of sets of hypercubes to represent the relations.
The basic principle is that applying a logical operator to one or more relations, amounts to applying this operator to their corresponding sets of hypercubes.
This leads to an implicit way to infer unknown relationships from known ones by application of the material (non)implication.

The only non-trivial unitary operator is the negation (NOT, $\neg$).
The negation of a relationship between $n$ sets is represented by the set of hypercubes of dimension $n$ that are not consistent with the original relationship.
This set can be constructed by creating the tautology of dimension $n$ and removing those hypercubes that were used to represent the original relationship (Algorithm \ref{algo:negation}).
This is not scalable, because the size of the tautology grows exponentially with the number of dimensions.
The negation should therefore be avoided, and thereby also its implicit use in binary operators.

Applying a binary operator to two relationships requires that the used hypercubes represent the same sets (Algorithm \ref{algo:binaryoperator}).
In general this can be achieved by adding sets to each relationship (through algorithm \ref{algo:addsetsimple}) until they represent relationships between identical sets.
However, in some cases it suffices to remove sets from the relationships (\refsec{sec:optimizations}).
Binary operators that are of particular importance for the deduction of a priori unknown relationships are:
\begin{itemize}
 \item Conjunction (AND, $\wedge$, $\setand$): Combines two relationships that are both known to hold.
The result of $P \wedge Q$ is a relationship represented by hypercubes that are consistent with both a hypercube in $P$ and one in $Q$.
 \item Disjunction (OR, $\vee$, $\setor$): Combines relationships of which it is known that at least one of them holds. 
The result of $P \vee Q$ is a relationship represented by hypercubes that are consistent with a hypercube in $P$ and/or one in $Q$.
 \item Material Implication ($\rightarrow$): Can be used to infer relations.
The result of $P\rightarrow Q$ is a relationship with hypercubes that are consistent with both $P$ and $Q$, together with those that are not consistent with $P$.
The relationship $P$ implies that relationship $Q$ holds when $P\rightarrow Q$ results in the tautology.
The material implication ($P \rightarrow Q$) can be implemented as $(\neg(P \wedge (\neg Q)))$, which requires the negation operator.
An implementation of the negation is not scalable; the material implication is therefore not suitable to prove whether unknown relations hold.
 \item Material Nonimplication ($\nrightarrow$): Can also be used to infer relationships.
The relation that is the result of the material nonimplication $(P \nrightarrow Q)$ is represented by the set of hypercubes that is consistent with $P$, but not with $Q$.
  This operation can be used to prove that relation $Q$ must hold given $P$, because $P$ implies $Q$ when the result of the operation is the contradiction.
  This operation is more suitable for implementation than the material implication, because it always results in a relation that is represented by less hypercubes than the original relations.
\end{itemize}

The logical operators can be used to prove that a specific relationship must hold by testing for entailment (Algorithm \ref{algo:entailment}).
First a list of relationships $(S_0, S_1, ...)$ is constructed, where each $S_i$ contains partial a priori knowledge about the sets.
The logical conjunction operator is subsequently applied to all these relationships, resulting in relationship $S$.
Finally, the nonimplication $S\nrightarrow R$ is applied, where $R$ is the relationship that needs to be proven.
Relationship $R$ must be valid if the result of the nonimplication is the contradiction.

\subsection{Optimizations}
\label{sec:optimizations}
The logical operators are discussed in the previous section in an intuitive but naive form that will lead to an unscalable implementation.
Firstly, adding a set to a relationship requires the creation of all hypercubes of a specific dimension.
This is not feasible for dimensions higher than about 4.
This can be avoided by not creating hypercubes with \booleantrue vertices that correspond to two \booleanfalse vertices in the original (Algorithm \ref{algo:addsetbetter}).
Secondly, enlarging the set of hypercubes in order to apply binary operators can sometimes be avoided entirely, in particular for conjunction and material nonimplication.

Adding sets to $P$ with the purpose of performing the conjunction $P\wedge Q$ can often be done without enlarging the number of hypercubes.
This is the case when for each hypercube of $P$ there is at most one more-dimensional hypercube that is consistent with both $P$ and $Q$.
Algorithm \ref{algo:conjunctionenhanced} shows how to verify this condition when only one of the sets in $Q$ is not in $P$.
The algorithm checks for a one-to-one correspondence between the hypercubes of $Q$ and the hypercubes of $Q$ with this extra set removed.
This correspondence, if existent, can be used to add the extra set to $P$ without enlarging the number of hypercubes.

The material nonimplication operator can sometimes be performed by \textit{removing} dimensions instead of adding them, because it tests for inconsistency (Algorithm \ref{algo:enhancedmaterialnonimplication}).
This is possible for the operation $P\nrightarrow Q$ when all the sets of $Q$ are also represented by $P$.
It is not necessary to add the extra dimensions of $P$ to $Q$ in order to test which hypercubes in $P$ are inconsistent with $Q$: the hypercubes of $Q$ essentially represent the tautology for these extra sets and it is not possible to be inconsistent with the tautology.
Instead, the extra dimensions can be removed from the hypercubes of $P$ to determine whether the originals are consistent with $Q$.

Furthermore, sets that are equal can be represented with the same dimension of the hypercubes.
This optimization would make the presentation of the algorithms more complicated without adding conceptual insights and is therefore not discussed in this paper.

\section[Astro-WISE]{\AW}
\label{sec:astrowise}
The presented method is used in the \AW information system to handle astronomical catalogs.
These catalogs contain information about astronomical objects and can therefore be seen as sets with these objects as elements.
Catalogs in \AW are primarily created either from images or by performing an operation on other catalogs; the mechanisms presented in this paper are only used for the latter kind.

\subsection{Objects and Dependency Graphs}
\AW uses an Object-Oriented data model in which science products are stored as class instantiations.
Every class forms a blueprint of how its instances should be processed to create the data from other objects.
Every object has \textit{persistent properties} that are stored in the database, which allows the object to be used across sessions and shared between scientists.
The persistent properties of an object include all the details of its processing: its \textit{dependencies}, and the values of any process parameters that can be set.
Different catalog classes are designed for different operations to create catalogs.

To create full \textit{data lineage}, the depencies of an object have their own dependencies.
This net of dependencies that links an object to the raw data is called a \textit{dependency graph} (\reffig{fig:scexamplebare}).
The algorithms presented in this paper are used for the automatic creation and manipulation of dependency graphs dealing with catalogs.

\subsection{Target Processing}
The heart of \AW is its request driven way of data handling, called \textit{target processing} \citep{Mwebaze:2009:ATU:1683300.1683752}.
In the traditional way of data handling, scientists start with a data set and perform operations until they reach their required end product.
Target processing turns this around: scientists request the desired end product directly---their \textit{target}---and the information system will create a dependency graph that ends with an object representing the requested data.
The information system can reuse existing objects, possibly created by other scientists.
Furthermore, it can autonomously create new objects, because the class definition forms a blueprint for new objects.

The data lineage allows any object to be processed at any time, because the object's class and persistent properties describe how this can be done.
This is taken to the extreme for catalog instances: catalogs can be created and stored without fully processing them, or without processing them at all \citep{pullingcatalogs}.
In other words, it is not required to create or store the contents of a catalog as a whole, achieving the scalability required to handle large catalogs.
Therefore, determining the contents of the catalogs should be possible without consulting the catalog data directly.

The information system can process catalogs partially by modification of dependency graphs.
This allows new catalogs to be created in their most general form to maximize their reusability for future requests.
At the same time this ensures that catalog data is only created when this is essential for the requested dataset.
Optimization of the dependency graphs requires the information system to know as much as possible about the contents of the catalogs in the graph \textit{before} they are processed.

\subsection{Algorithm Specifics}
The presented method determines whether a desired relation holds by combining information about known relationships between catalogs.
The catalog classes for \AW are designed to maximize this a priori knowledge.
In particular, every catalog class corresponds to a specific operation to derive catalog data from other objects.
Many of these correspond to relational operators \citep{Codd:1970:RMD:362384.362685}.
Each catalog class allows only a specific set of relationships between the sets of sources of a catalog and its dependencies.

Every catalog instance has partial knowledge about its relationship with its dependencies: it knows which relations are permitted by its class, not which of those actually holds.
A priori this is the only available information.
The presented mechanism is used to acquire knowledge that requires combining this local information.

In this astronomical setting, sets correspond to astronomical catalogs, and the elements are astronomical objects.
This background puts several constraints on the use of the algorithm:
\begin{itemize}
 \item All catalogs by design have one of the following relationships with their dependencies, with in brackets the number of corresponding hypercubes (\reftab{tab:hypercubeslowd}): equality (4), subset (4), intersection (16) or union (16). However, they can have any relationship with catalogs that are not their direct dependency.
 \item The following relations are the most important in checking which relations hold between sets: non-emptiness (2), equality (4), superset (4).
 \item A relation where all objects are within a set can never hold.
\end{itemize}
Most of the relations that are enforced by the catalog classes are shown in \reftab{tab:hypercubeslowd}.
In \refsec{sec:example2} the method is applied to a simplified \AW use case.

\subsection{Scalability}
The major factor in the scalability of the method is the size of the set of hypercubes used to represent relationships.
Adding a dimension to a hypercube, will result in $3^t$ new hypercubes, where $t$ is the number of \booleantrue cells in the original hypercube.
The size of the hypercubes themselves is less of an issue: these scale with $2^{n}$, while the number of possible hypercubes scales with $2^{2^n}$ where $n$ is the number of sets.
The number of possible hypercubes can grow very rapidly with the number of sets when little is known about their relationship.
However, this is not necessarily problematic for application in \AW:
\begin{itemize}
 \item Many catalogs represent the exact same objects.
It is not required to add a new dimension to the hypercubes when adding a set that is known to be equal to one of the other sets: the set can be associated with an existing dimension.
 \item Sets that are different can still have a relation that is quantified by a low number of \booleantrue cells. 
For example, sets that are a subset of another set occur often and require only one extra \booleantrue cell.
Furthermore, some sets, e.g. those that are the intersection of sets already in a relation, can be added without increasing the number of \booleantrue cells at all (Algorithm \ref{algo:conjunctionenhanced}).
The relations that require the most \booleantrue cells, such as disjoint or partially overlapping sets, are rare, because comparisons are done on catalogs that are connected through data lineage.
 \item Some relations are very unlikely to occur at all. For example there will always be objects not in any set.
\end{itemize}
Nonetheless, the set of hypercubes can become large for large dependency graphs of catalogs.
However, in most cases the number of hypercubes can be limited:
\begin{itemize}
 \item External knowledge---with respect to this algorithm---can be used explicitly.
 For example it can often be determined whether a catalog is empty or whether two catalogs are disjoint.
 \item Any knowledge about the relationships that is obtained, through the algorithm or otherwise, can be stored for future use.
 \item The sets of hypercubes are created by traversing the dependency graphs of catalogs.
 The most interesting relationships in a dependency graph are those between the begin and end points.
 Dimensions that correspond to catalogs in the middle of a dependency graph might be removed when this has little or no influence on the relationships between the catalogs at the edges.
\end{itemize}
This combination of factors ensures that the algorithm is sufficiently scalable to meet the requirements for use in \AW.

\setlength{\tabcolsep}{1pt}
\begin{table}
\centering
 \begin{tabular}{|c|c|c|c|c|c|c|c|c|c|c|c|c|c|}
  \hline
    \incvenn{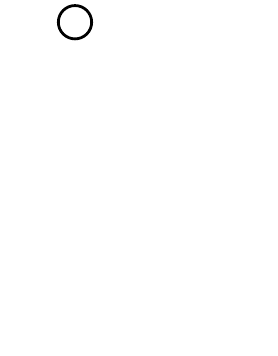} &
    \multicolumn{11}{c|}{\incvenn{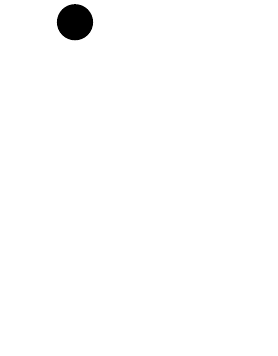}} &
    0D\\
  \hline
  \hline
    \incvenn{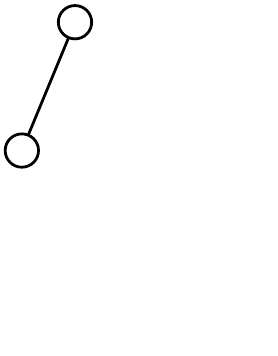} &
    \multicolumn{1}{c|}{\incvenn{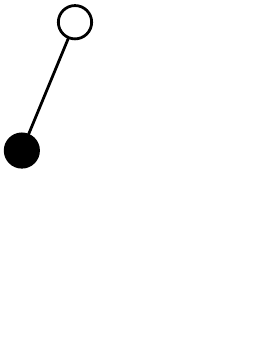}} &
    \multicolumn{3}{c|}{\incvenn{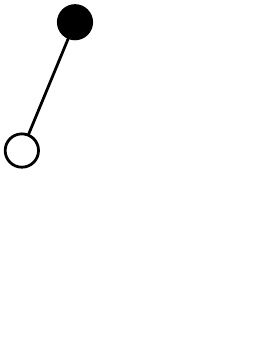}} &
    \multicolumn{7}{c|}{\incvenn{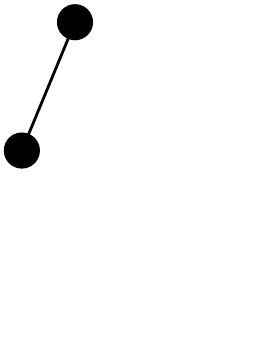}} &
    1D \\
  \hline
    \checkmark &
    \multicolumn{1}{c|}{} &
    \multicolumn{3}{c|}{\checkmark} &
    \multicolumn{7}{c|}{} &
    empty \\
  \hline
  \hline
    \incvenn{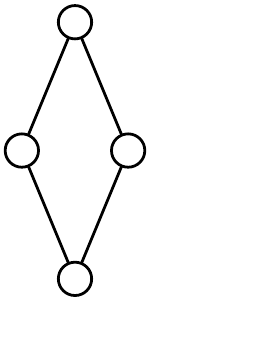} &
    \multicolumn{1}{c|}{(3)} &
    \multicolumn{1}{c|}{\incvenn{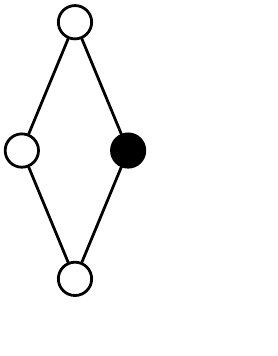}} &
    \multicolumn{1}{c|}{\incvenn{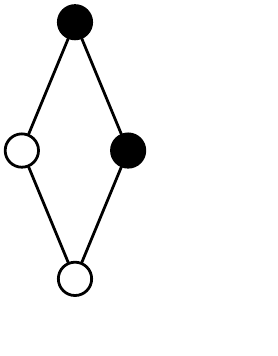}} &
    \multicolumn{1}{c|}{\incvenn{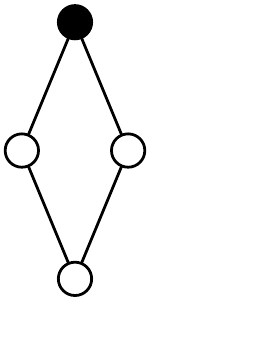}} &
    \hspace{0.1cm}(3)\hspace{0.1cm} &
    \multicolumn{1}{c|}{\incvenn{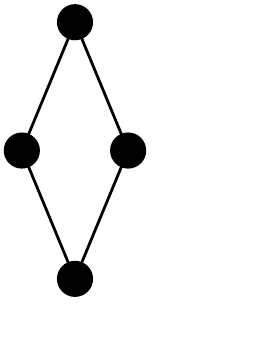}} &
    \multicolumn{1}{c|}{\incvenn{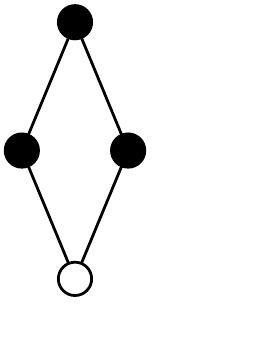}} &
    \multicolumn{1}{c|}{\incvenn{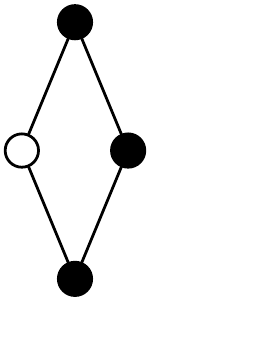}} &
    \multicolumn{1}{c|}{\incvenn{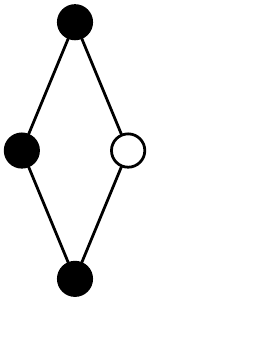}} &
    \multicolumn{1}{c|}{\incvenn{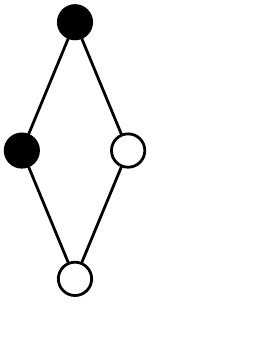}} &
    \multicolumn{1}{c|}{\incvenn{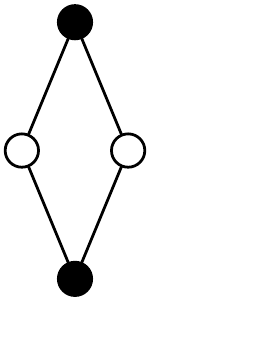}} &
    2D \\
  \hline
    \checkmark & \checkmark  & &  & \checkmark  & &  &  & & & & \checkmark & equality\\
      & \checkmark \hspace{-0.1cm}\checkmark & &  &   &  &  &  &  &\checkmark &\checkmark &   & subset\\
      & & \checkmark & \checkmark &   & \checkmark &  &  &\checkmark  && &   & superset\\
  \hline
  \hline
    \incvenn{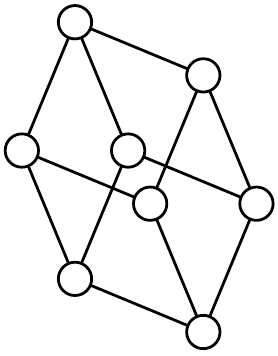} &
    \multicolumn{1}{c|}{(3)} &
    \multicolumn{1}{c|}{\incvenn{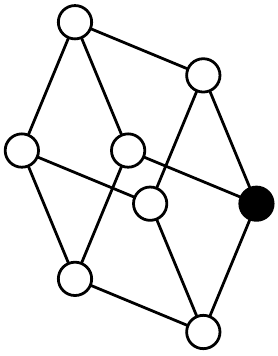}} &
    \multicolumn{1}{c|}{\incvenn{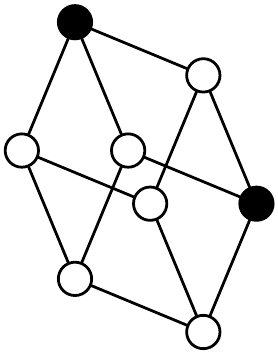}} &
    \multicolumn{1}{c|}{\incvenn{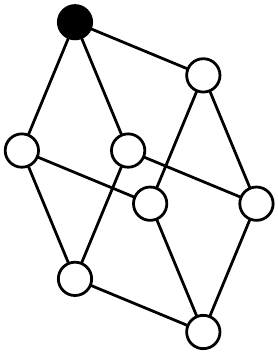}} &
    \multicolumn{1}{c|}{(3)} &
    \multicolumn{1}{c|}{\incvenn{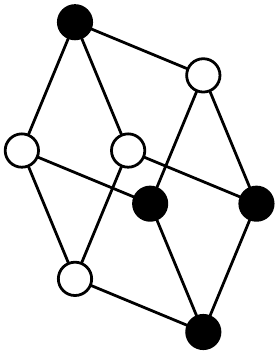}} &
    \multicolumn{1}{c|}{\incvenn{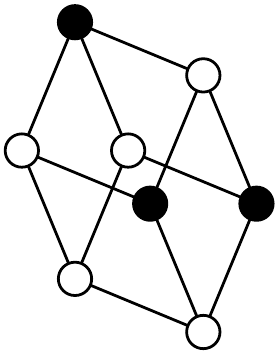}} &
    \multicolumn{1}{c|}{\incvenn{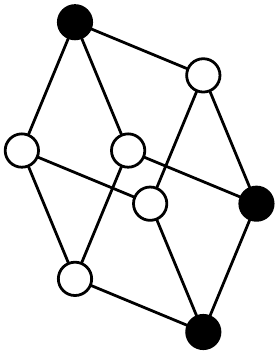}} &
    \multicolumn{1}{c|}{\incvenn{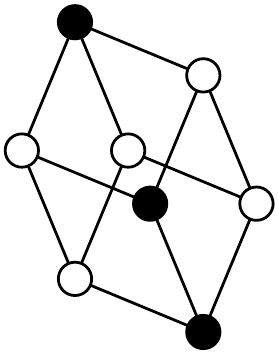}} &
    \multicolumn{1}{c|}{\incvenn{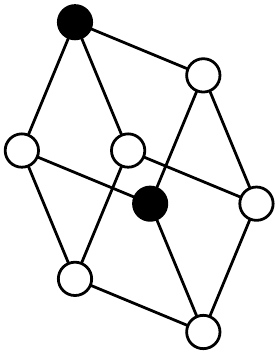}} &
    \multicolumn{1}{c|}{\incvenn{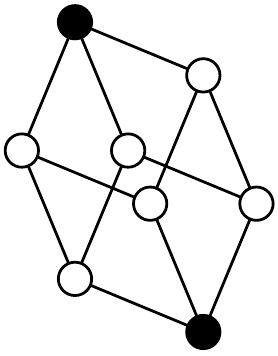}} &
    3D union\\
  \hline
    \incvenn{00000000} &
    \multicolumn{1}{c|}{(3)} &
    \multicolumn{1}{c|}{\incvenn{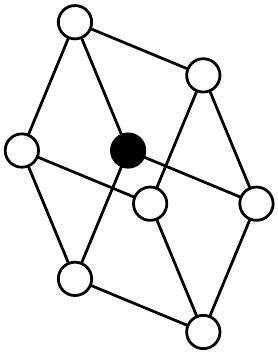}} &
    \multicolumn{1}{c|}{\incvenn{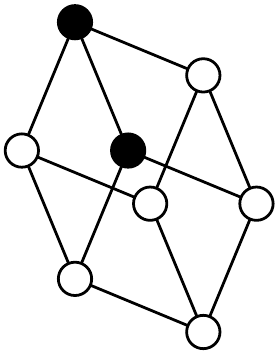}} &
    \multicolumn{1}{c|}{\incvenn{10000000}} &
    \multicolumn{1}{c|}{(3)} &
    \multicolumn{1}{c|}{\incvenn{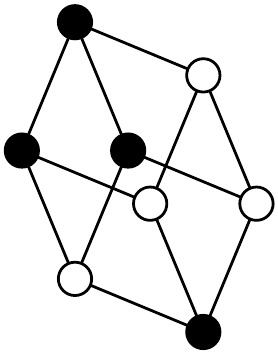}} &
    \multicolumn{1}{c|}{\incvenn{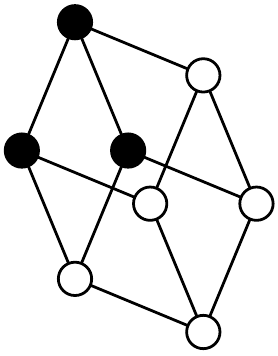}} &
    \multicolumn{1}{c|}{\incvenn{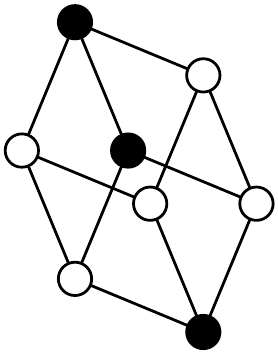}} &
    \multicolumn{1}{c|}{\incvenn{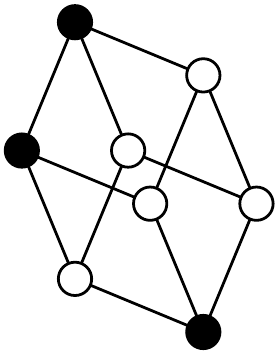}} &
    \multicolumn{1}{c|}{\incvenn{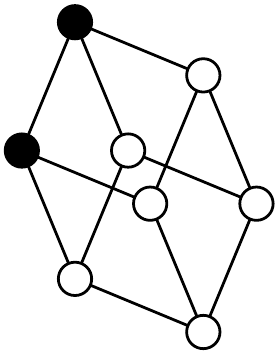}} &
    \multicolumn{1}{c|}{\incvenn{10000001}} &
    intersec. \\
  \hline
 \end{tabular}
 \caption{Examples of hypercube representations of low dimensional relations.
Relations that are not directly relevant to our astronomical application are omitted.
They are replaced with a number in parenthesis that indicates the number of missing hypercubes.
The top vertex of each hypercube represents objects not in any of the sets under consideration, the bottom vertex objects in all the sets.
A solid circle is used for \booleantrue values and an open circle for \booleanfalse.
The hypercubes are ordered hierarchically: a relation in a lower cell implies the relations of lower dimension in the cells above it.
The check marks at one and two dimensions indicate that a set of these hypercubes represents the relation mentioned in the last column.
Check marks below the numbers in parenthesis indicate the number of omitted hypercubes that are part of this set.
The \textit{superset} and \textit{subset} labels in the 2D rows refer to the extra dimension with respect to the one already present in the 1D row.
Furthermore they are \textit{strict}: an equality is not considered a subset.
Of the 256 three dimensional relations, only those where the third dimension corresponds to the union or intersection of the first two are shown.
}
\label{tab:hypercubeslowd}
\end{table}

\section{Pseudocode}
\label{sec:pseudocode}
The pseudocode for the algorithms mentioned above is presented.
Every relationship $P$ is assumed to be represented with a set of hypercubes $H_p$ and a set of labels $\Lambda_p$.
These labels identify the dimensions of the hypercubes with the sets considered by the relationship.
The administration of these labels is trivial and is therefore only discussed when relevant for handling the hypercubes.

Dimensions of the hypercubes are denoted with $\nu$'s. A specific vertex or cell in a hypercube $h_p\epsilon H_p$ is denoted with $h_p(\nu_1, \nu_2, ..., \nu_n)$, where each $\nu_i$ can have a value of $0$ or $1$.
It is assumed that the dimensions of the hypercubes are in the same order when they are compared.
A transposition of the dimensions suffices to accomplish this when necessary.

\begin{algorithm}[hbtp]
\caption{Removing a set from a relationship}
\label{algo:removeset}
\begin{algorithmic}[1]
\REQUIRE $H_o =$ set of hypercubes of the original relationship
\REQUIRE $d =$ dimension corresponding to the to-be-removed set.
\ENSURE $H_r =$ set of all hypercubes consistent with $H_o$ without dimension $\nu_d$
\STATE $n \leftarrow$ dimension of the hypercubes in $H_o$
\STATE $H_r \leftarrow$ empty set of hypercubes
\FORALL{hypercubes $h_o$ in $H_o$}
  \STATE $h_r \leftarrow$ hypercube of dimension $n-1$ with all values set to \booleanfalse
  \FORALL{cells $c_r = h_r(\nu_1,...,\nu_{d-1},\nu_{d+1},...,\nu_{n})$ in $h_r$}
    \STATE $c_p \leftarrow h_o(\nu_1,...,\nu_{d-1},0,\nu_{d+1},...,\nu_{n})$
    \STATE $c_q \leftarrow h_o(\nu_1,...,\nu_{d-1},1,\nu_{d+1},...,\nu_{n})$
    \STATE $h_r(\nu_1,...,\nu_{d-1},\nu_{d+1},...,\nu{n}) \leftarrow c_p \vee c_q$
  \ENDFOR
  \STATE $H_r \leftarrow H_r \setor \mathrm{set}(h_r)$
\ENDFOR
\end{algorithmic}
\end{algorithm}

\begin{algorithm}[hbtp]
\caption{Adding a set to a relationship in a naive way}
\label{algo:addsetsimple}
\begin{algorithmic}[1]
\REQUIRE $H_o =$ set of hypercubes of the original relationship
\ENSURE $H_r =$ set of all hypercubes consistent with $H_o$ with one extra dimension
\STATE $n \leftarrow$ dimension of the hypercubes in $H_o$
\STATE $H_r \leftarrow$ empty set of hypercubes
\STATE $H_t \leftarrow$ set of all $2^{\left(2^{n+1}\right)}$ distinct hypercubes of dimension $n+1$
\FORALL{hypercubes $h_t$ in $H_t$}
  \STATE $h_u \leftarrow h_t$ with dimension $\nu_{n+1}$ removed (Algorithm \ref{algo:removeset})
  \IF{$h_u$ in $H_o$}
    \STATE $H_r \leftarrow H_r \setor \mathrm{set}(h_t)$
  \ENDIF
\ENDFOR
\end{algorithmic}
\end{algorithm}

\begin{algorithm}[hbtp]
\caption{Applying the Negation operator: $r=\neg q$}
\label{algo:negation}
\begin{algorithmic}[1]
\REQUIRE $H_p =$ set of hypercubes of the original relationship
\ENSURE $H_r =$ set of all hypercubes of the same dimension that are not consistent with $H_p$
\STATE $n \leftarrow$ dimension of the hypercubes in $H_p$
\STATE $H_r \leftarrow$ empty set of hypercubes
\STATE $H_t \leftarrow$ set of all $2^{\left(2^n\right)}$ distinct hypercubes of dimension $n$
\FORALL{hypercubes $h_t$ in $H_t$}
  \IF{not $h_t$ in $H_p$}
    \STATE $H_r \leftarrow H_r \setor \mathrm{set}(h_t)$
  \ENDIF
\ENDFOR
\end{algorithmic}
\end{algorithm}

\begin{algorithm}[hbtp]
 \caption{Applying any binary operator: $r=p\circ q$}
 \label{algo:binaryoperator}
 \begin{algorithmic}[1]
  \REQUIRE $H_p =$ set of hypercubes of the first relationship
  \REQUIRE $\Lambda_p =$ list of labels that correlates sets to the dimensions of $H_p$
  \REQUIRE $H_q =$ set of hypercubes of the second relationship
  \REQUIRE $\Lambda_q =$ list of labels that correlates sets to the dimensions of $H_q$
  \REQUIRE $\circ =$ the logical operation to be applied to the relationship
  \ENSURE $H_r =$ set of hypercubes that is consistent with both $H_p$ and $H_q$
  \FORALL{labels $\lambda_p$ in $\Lambda_p$ not in $\Lambda_q$}
    \STATE $H_v \leftarrow H_q$ with dimension $\lambda_p$ added (Algorithm \ref{algo:addsetsimple}, \ref{algo:addsetbetter})
  \ENDFOR
  \FORALL{labels $\lambda_q$ in $\Lambda_q$ not in $\Lambda_p$}
    \STATE $H_u \leftarrow H_p$ with dimension $\lambda_q$ added (Algorithm \ref{algo:addsetsimple}, \ref{algo:addsetbetter})
  \ENDFOR
  \STATE $H_r \leftarrow H_u \circ H_v$
 \end{algorithmic}
\end{algorithm}

\begin{algorithm}[hbtp]
 \caption{Testing for Entailment}
\label{algo:entailment}
\begin{algorithmic}[1]
 \REQUIRE $P_i = $ set of relationships representing the a priori knowledge, $i=1,2,..$.
 \REQUIRE $Q = $ a relation for which it is unknown whether it holds
 \ENSURE $r = $ \booleantrue when $P_i$ entails $Q$, \booleanfalse otherwise
 \STATE $P \leftarrow (P_0 \wedge (P_1 \wedge ... ))$ 
 \STATE $R \leftarrow (P \nrightarrow Q)$
 \IF{$R ==$ Contradiction}
   \STATE $r \leftarrow$ \booleantrue
 \ELSE
   \STATE $r \leftarrow$ \booleanfalse
 \ENDIF
\end{algorithmic}
\end{algorithm}

\begin{algorithm}[hbtp]
\caption{Improved way of adding a set to a relation}
\label{algo:addsetbetter}
\begin{algorithmic}[1]
\REQUIRE $H_o =$ set of hypercubes of the original relation
\ENSURE $H_r =$ set of all hypercubes consistent with $H_o$ with one extra dimension
\STATE $n \leftarrow$ dimension of the hypercubes in $H_o$
\STATE $H_r \leftarrow$ empty set of hypercubes
\FORALL{hypercubes $h_o$ in $H_o$}
\STATE $h_t \leftarrow$ hypercube of dimension $n+1$ with all values set to \booleanfalse
\FORALL{cells $c_o = h_o(\nu_1,\nu_2,...,\nu_n)$ in $h_o$}
\STATE $h_t(\nu_1,\nu_2,...,\nu_n,0) \leftarrow c_o$
\STATE $h_t(\nu_1,\nu_2,...,\nu_n,1) \leftarrow c_o$
\ENDFOR
\STATE $H_t \leftarrow \mathrm{set}(h_t)$
\FORALL{cells $c_o = h_o(\nu_1,\nu_2,...,\nu_n)$ in $h_o$}
\IF{$c_o==\mathtt{True}$}
\FORALL{hypercubes $h_t$ in $H_t$}
\STATE $h_u \leftarrow h_t$
\STATE $h_u(\nu_1,\nu_2,...,\nu_n,0) \leftarrow \mathtt{False}$
\STATE $h_v \leftarrow h_t$
\STATE $h_v(\nu_1,\nu_2,...,\nu_n,1) \leftarrow \mathtt{False}$
\STATE $H_t \leftarrow H_t \setor \mathrm{set}(h_u, h_v)$
\ENDFOR
\ENDIF
\ENDFOR
\STATE $H_r \leftarrow H_r \setor H_t$
\ENDFOR
\end{algorithmic}
\end{algorithm}

\begin{algorithm}[hbtp]
\caption{Enhanced Conjunction: $r=p\wedge q$}
\label{algo:conjunctionenhanced}
\begin{algorithmic}[1]
\REQUIRE $H_p =$ set of hypercubes of dimension $n_p$ of the original relation
\REQUIRE $H_q =$ set of hypercubes of dimension $n_q$ where only the last dimension $\nu_{n_q}$ is not present in $H_p$ and the first dimensions $\nu_{1}$ to $\nu_{n_q-1}$ correspond to the first dimensions of $H_p$
\ENSURE $H_r =$ set of hypercubes of dimension $n_p+1$ that is consistent with both $H_p$ and $H_q$
\STATE $H_t \leftarrow H_q$ with dimension $n_q$ removed (Algorithm \ref{algo:removeset})
\IF{length($H_t$) == length($H_q$)}
  \STATE $h_t \leftarrow$ hypercube of dimension $n_q$ with all values set to \booleanfalse
  \FORALL{hypercubes $h_q$ in $H_q$}
    \FORALL{cells $c_t = h_t(\nu_1,\nu_2,...,\nu_{n_q})$ in $h_t$}
      \STATE $h_t(\nu_1,\nu_2,...,\nu_{n_q}) \leftarrow c_t \vee h_q(\nu_1,\nu_2,...,\nu_{n_q})$
    \ENDFOR
  \ENDFOR
  \STATE $H_r \leftarrow$ empty set of hypercubes
  \FORALL{hypercubes $h_p$ in $H_p$}
    \STATE $h_r \leftarrow$ hypercube of dimension $n_p+1$ with all values set to \booleanfalse
    \FORALL{cells $c_p = h_p(\nu_1,\nu_2,...,\nu_{n_p})$ in $h_p$}
      \IF{$c_p == True$}
        \IF{$h_t(\nu_1,\nu_2,...,\nu_{n_q-1},0)==True$}
          \STATE $h_r(\nu_1,\nu_2,...,\nu_{n_p}, 0) \leftarrow True$
        \ELSE
          \STATE $h_r(\nu_1,\nu_2,...,\nu_{n_p}, 1) \leftarrow True$
        \ENDIF
      \ENDIF
    \ENDFOR
    \STATE $h_m \leftarrow h_r$ with dimension $n_p+1$ removed
    \IF{$h_m$ in $H_p$}
      \STATE $H_r \leftarrow H_r \setor \mathrm{set}(h_r)$
    \ENDIF
  \ENDFOR
\ELSE
  \STATE Create $H_r$ through the regular conjunction with algorithm \ref{algo:binaryoperator}.
\ENDIF
\end{algorithmic}
\end{algorithm}

\begin{algorithm}[hbtp]
 \caption{Enhanced Material Nonimplication: $r=p\nrightarrow q$}
 \label{algo:enhancedmaterialnonimplication}
 \begin{algorithmic}
  \REQUIRE $H_p =$ set of hypercubes of the first original relation
  \REQUIRE $\Lambda_p =$ list of labels that correlates sets to the dimensions of $H_p$
  \REQUIRE $H_q =$ set of hypercubes of the second original relation
  \REQUIRE $\Lambda_q =$ list of labels that correlates sets to the dimensions of $H_q$
  \REQUIRE $\mathrm{set}(\Lambda_q) \subseteq \mathrm{set}(\Lambda_p)$
  \ENSURE $H_r =$ set of hypercubes that is consistent with $H_p$ but not with $H_q$
  \STATE $H_r \leftarrow$ empty set of hypercubes
  \FORALL{hypercubes $h_p$ in $H_p$}
    \STATE $h_t \leftarrow h_p$
    \FORALL{$\lambda_t$ in $\Lambda_p$ not in $\Lambda_q$}
      \STATE $h_t \leftarrow h_t$ with dimension corresponding to $\lambda_t$ removed (Algorithm \ref{algo:removeset})
    \ENDFOR
    \IF{$h_t$ in $H_q$}
      \STATE $H_r \leftarrow H_r \setor \mathrm{set}(h_p)$
    \ENDIF
  \ENDFOR
 \end{algorithmic}
\end{algorithm}

\section{Example}
\label{sec:example2}
\begin{figure}[ht!]
 \centering
 \includegraphics[width=0.4\linewidth]{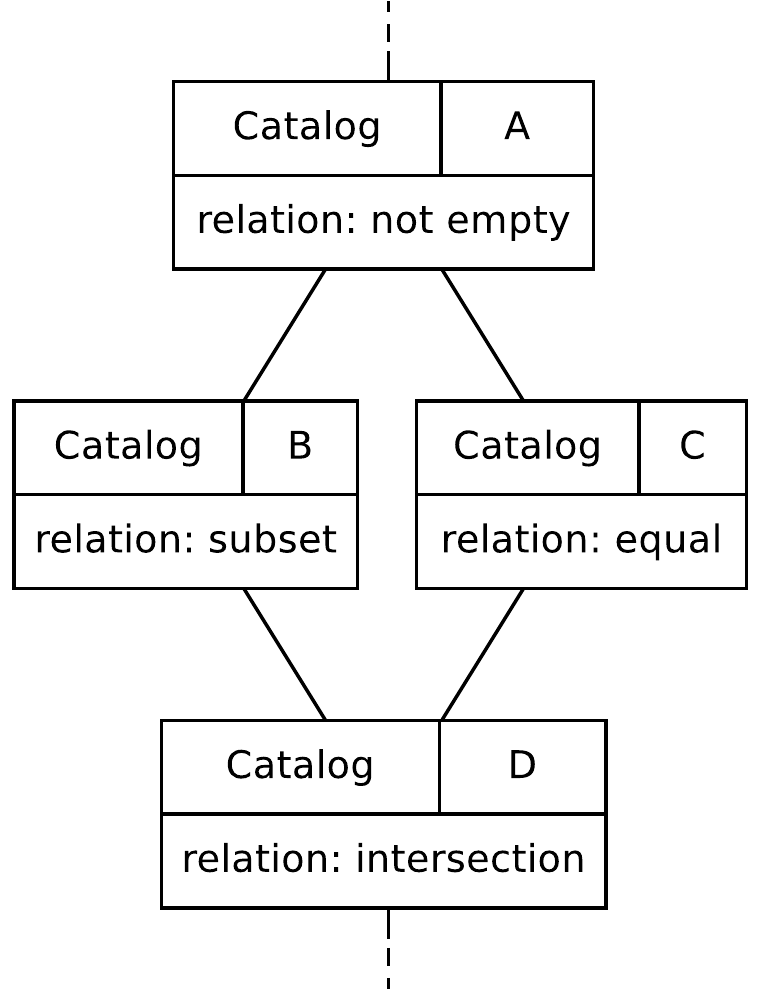}
 \caption{A simplified part of a dependency graph in \AW. 
Catalog $A$ contains a known set of sources.
Catalog $B$ represents an yet unknown subset of the sources of $A$.
Catalog $C$ represents the same sources as $A$ with a different set of attributes.
Catalog $D$ represents an intersection of the sources of $B$ and $C$ with the attributes of both $B$ and $C$.
}
 \label{fig:scexamplebare}
\end{figure}

\begin{figure}[ht!]
\centering
\includegraphics[width=0.9\linewidth]{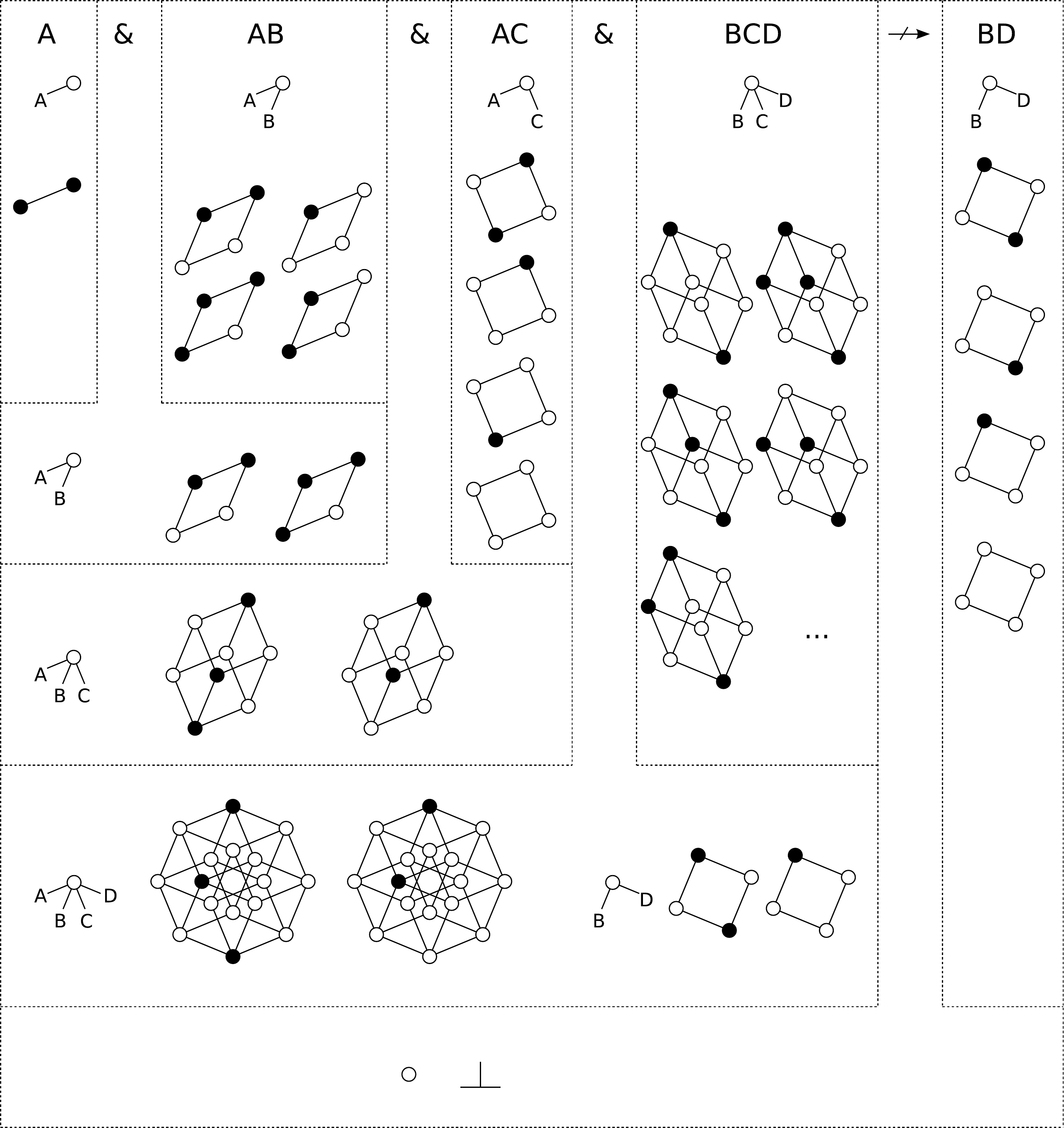}
\caption{The determination that catalogs $B$ and $D$ of \refsec{sec:example2} (\reffig{fig:scexamplebare}) represent the same astronomical objects.
The hypercubes are projected such that a specific set is always oriented in the same direction.
A solid circle represents a \booleantrue value and an open circle a \booleanfalse one.
The circle at the top represents the first cell in all dimensions.
The four dimensional hypercubes of the relationship between $A$, $B$, $C$ and $D$ are also shown in reduced form as a relationship between $B$ and $D$.
$B$ must be equal to $D$ because both hypercubes representing the relationship between $A$, $B$, $C$ and $D$ are consistent with this equality.
}
\label{fig:example2man}
\end{figure}

The method is demonstrated with a simplified application in \AW.
We will use the terms \textit{source} for an astronomical object in a catalog, that is, an element in a set.
Furthermore we use the term \textit{attribute} for a quantified physical property of that object, for example its mass.
\Reffig{fig:scexamplebare} shows a simplified part of a dependency graph consisting of four catalogs:
\begin{itemize}
 \item Catalog $A$ is the base catalog from which the others are derived and contains a finite, known, set of sources.
 The catalog does not contain all the sources in the Universe.
 \item Catalog $B$ represents a subset of the sources of $A$.
The selection criterion is known, but unevaluated.
The contents of $B$ is therefore unknown, and it might even be empty.
 \item Catalog $C$ represents new attributes of the sources in catalog $A$.
That is, the attributes are not in catalog $A$ and have to be derived.
 The values of these attributes do not have to be calculated or stored in order to create the dependency graph.
 Catalog $C$ represents the same sources as catalog $A$.
 \item Catalog $D$ combines the attributes of catalogs $B$ and $C$ and represents an intersection of their sources.
 The precise contents of this catalog is unknown at its creation, because the selection criterion of $B$ is not yet evaluated and the attributes of $C$ are not yet calculated.
\end{itemize}
Such a dependency tree would have been created automatically by requesting the attributes from $A$ and $C$ for the sources specified in $B$.
The information system will attempt to process this dependency graph in an optimal way.
In this case, it will try to limit the processing of catalog $C$ to those sources that are required to process $D$.
A priori, the only available information about $D$ is the local knowledge that $D$ has about its relationship with $B$ and $C$.
The algorithm is applied to determine that set $D$ represents the exact same sources as $B$.
The following steps are performed, visualized in \reffig{fig:example2man}:
\begin{itemize}
 \item All the hypercubes consistent with the local information are created as relationships $A$ (nonempty), $AB$ (subset), $AC$ (equality) and $BCD$ (intersection).
 \item The conjunction operator is subsequently applied on these relationships.
 Dimensions are added to the hypercubes when necessary.
The result is a four dimensional relationship between $A$, $B$, $C$ and $D$.
 \item Relationship $BD$ is created, representing an equality between $B$ and $D$.
It is a priori unknown whether this relation holds.
 \item The material nonimplication operator is applied to relation $ABCD$ and $BD$, resulting in the contradiction (\refsec{sec:relasset}). That is, there are no possible relationships between $A$, $B$, $C$ and $D$---given the a priori knowledge---in which $B$ and $D$ are not equal.
Therefore, $B$ and $D$ must represent the same sources.
\end{itemize}
The information system will optimize the dependency graph using the knowledge that the sources in $B$ and $D$ are equal.
In particular, it will evaluate the selection criterion in $B$ and only calculate the attributes of $C$ for those sources.
The conclusion that $B$ is equal to $D$ is reached without having to consult any catalog data, which was necessary because the catalog data had not yet been created.

\section{Conclusions}
\label{sec:conclusions}
A novel mechanism for inferring relationships between sets is discussed.
It is shown that the use of sets of hypercubes to represent relationships leads to a natural way of inferring a priori unknown relations:
deduction is performed by combining incomplete knowledge through the application of logical operators.
Algorithms that are suitable for a scalable implementation are presented, including pseudocode.

The novel aspects of the method were demonstrated by its use in \AW, where the sets correspond to catalogs and the elements to astronomical objects.
Catalogs can be stored and used in \AW without their content being evaluated.
The method is used to acquire knowledge about their contents without requiring direct access to the catalog data.
This has lead to design choices in the way catalogs are handled in \AW: catalogs are created such that the knowledge about their relationships is maximized.

An automated way to infer relations between catalogs is essential for the request driven way of processing in information systems such as \AW.
The presented algorithms form an excellent method to accomplish this.
The method is generic enough to be implemented in any programming language and can be used by any information system.

\begin{acknowledgements}
This research is part of the project ``Astrovis'', research program STARE
(STAR E-Science), funded by the Dutch National Science Foundation (NWO),
project no.\ 643.200.501.
\AW is an on-going project which started from a FP5 RTD programme funded by the EC Action ``Enhancing Access to Research Infrastructures''.
The authors thank the anonymous reviewers for their insightful comments, which led to a higher quality and better structure of the paper and indirectly to new research directions.
\end{acknowledgements}


\begin{thebibliography}{}
\bibitem[Buddelmeijer et al.(2012)]{pullingcatalogs}Buddelmeijer,~H., Valentijn,~E.A.: Automatic optimized discovery, creation and processing of astronomical catalogs. Exp. Astron. (2011), doi:10.1007/s10686-011-9272-z
\bibitem[Chaudhuri(1998)]{Chaudhuri:1998:OQO:275487.275492}Chaudhuri~S. (1998) An overview of query optimization in relational systems. Proceedings of the seventeenth ACM SIGACT-SIGMOD-SIGART symposium on Principles of database systems. PODS '98 34-43 
\bibitem[Clarke(1994)]{Clarke199497short}Clarke~M.~C. (1994) Visualizing boolean operations on a hypercube. Mathematical and Computer Modelling 20(9):97--103
\bibitem[Codd(1970)]{Codd:1970:RMD:362384.362685}Codd, E.F.: A relational model of data for large shared data banks. Commun. ACM 13, 377–387 (1970)
\bibitem[Karnaugh(1953)]{k:map:1953}Karnaugh~M. (1953) The Map Method for Synthesis of Combinational Logic Circuits. Trans. AIEE pt. I 72(9):593--599
\bibitem[Mwebaze et al.(2009)]{Mwebaze:2009:ATU:1683300.1683752}Mwebaze, J., Boxhoorn, D., Valentijn, E.A.: Astro-wise: Tracing and using lineage for scientific data processing. In: Proceedings of the 2009 International Conference on Network-Based Information Systems, NBIS ’09, pp. 475–480. IEEE Computer Society, Washington, DC, USA (2009)
\bibitem[Sutcliff(2009)]{SutcliffTPTP}Sutcliffe~G. (2009) The TPTP Problem Library and Associated Infrastructure. The FOF and CNF Parts, v3.5.0. Journal of Automated Reasoning 43(4):337-362
\bibitem[Venn(1880)]{Venn1880}Venn~J.: On the diagrammatic and mechanical representation of propositions and reasonings. Phil. Mag., Series 5 10(59), 1-18 (1880)

    \end{thebibliography}
\end{document}